\documentclass[pdflatex,sn-mathphys-num]{sn-jnl}

\usepackage{orcidlink}
\usepackage{graphicx}%
\usepackage{multirow}%
\usepackage{amsmath,amssymb,amsfonts}%
\usepackage{amsthm}%
\usepackage{mathrsfs}%
\usepackage[title]{appendix}%
\usepackage{xcolor}%
\usepackage{textcomp}%
\usepackage{manyfoot}%
\usepackage{booktabs}%
\usepackage{algorithm}%
\usepackage{algorithmicx}%
\usepackage{algpseudocode}%
\usepackage{listings}%
\usepackage{amsmath}

\usepackage{natbib}
\usepackage{xparse}

\newcommand{\bi}{\begin{itemize}}

\usepackage{amsmath}
\usepackage{array}

\newcommand{\be}{\begin{equation}\begin{array}{lllllllllllll}}

\newcommand{\beno}{\begin{equation}\nonumber\begin{array}{lllllllllllll}}

\newcommand{\ee}{\end{array}\end{equation}}

\newcommand{\ei}{\end{itemize}}

\newcommand{\hide}[1]{}
\newcommand{\unitset}{\mathscr{P}_n}
\newcommand{\dyadset}{{\mathscr{D}_n}}
\newcommand{\timeset}{{\mathscr{T}_T}}
\newcommand{\dt}{\,\mathrm{d}t}
\newcommand{\dr}{\,\mathrm{d}r}
\newcommand{\ds}{\,\mathrm{d}s}
\newcommand{\dM}{\,\mathrm{d}M}
\newcommand{\dsum}{\displaystyle\sum\limits}

\newcommand{\dint}{\displaystyle\int\limits}
\newcommand{\dprod}{\displaystyle\prod\limits}
\newcommand{\dd}{\text{d}}

\usepackage{verbatim}
\usepackage{hyperref,bbm}
\usepackage{textcomp}
\usepackage{amsmath,mathtools,latexsym,mathtools}
\usepackage{bm}
\usepackage{amsfonts,amsmath}
\usepackage[mathscr]{eucal}
\usepackage{amssymb}
\usepackage{graphicx}

\def\bse{\begin{eqnarray*}}
\def\ese{\end{eqnarray*}}
\def\bc{\[\begin{array}{ccccccc}}
\def\ec{\end{array}\]}
\def\bsq{\begin{equation*}}
\def\esq{\end{equation*}}
\def\bq{\begin{equation}}
\def\eq{\end{equation}}

\let\oldeqref\eqref

\renewcommand{\eqref}[2][]{%
  \ifstrequal{#1}{s}%
    {\oldeqref{#2}}
    {Equation~\oldeqref{#2}}
}

\NewDocumentCommand{\hist}{g}{%
  \IfNoValueTF{#1}
    {\mathscr{F}_{t^-}}    
    {\mathscr{F}_{#1}} 
}

\newcommand{\timenow}{t^-}

\usepackage{bm}
\usepackage{amsmath}

\usepackage{amsmath}

\usepackage{subcaption}

\newcommand{\alertC}[1]{\textcolor{red}{\bf{CF: #1}}}
\newcommand{\alertA}[1]{\textcolor{blue}{\bf{AFK: #1}}}

\newcommand{\sampsize}{\mathcal{S}}
\newcommand{\para}{\bm{\theta}}
\newcommand{\parspace}{\Theta}
\newcommand{\truepara}{\bm{\theta}^*}
\newcommand{\paraneighbour}{\bm{\Theta}^*}
\newcommand{\paraspace}{\bm{\Theta}}
\newcommand{\semipara}{\bm{\eta}}

\newcommand{\MLE}{\hat{\bm{\theta}}_\sampsize}
\newcommand{\covs}{\bm{s}}
\newcommand{\semicovs}{\bm{r}}

\newcommand{\Var}{\textrm{Var}}
\usepackage{natbib}

\newcommand{\IR}{\mathbb{R}}
\newcommand{\IN}{\mathbb{N}}
\newcommand{\IP}{\mathbb{P}}
\newcommand{\IE}{\mathbb{E}}

\newcommand{\Ind}{\mathbbm{1}}
\theoremstyle{definition}
\newtheorem{theorem}{Theorem}[section]
\newtheorem{lemma}[theorem]{Lemma}

\newtheorem{remark}[theorem]{Remark}
\newtheorem{example}[theorem]{Example}

\newtheorem{assumptionA}{Assumption}  
\newtheorem{assumptionR}{Assumption}  

\theoremstyle{thmstyleone}%
\theoremstyle{thmstyletwo}%

\theoremstyle{thmstylethree}%

\begin{document}

\title[A Counting Process View of Relational Event Model]{A Counting Process View of Relational Event Models: Practical Asymptotics}


\author*[1]{\fnm{Cornelius} \sur{Fritz} \orcidlink{0000-0002-7781-223X}}\email{c.fritz@tcd.ie} 
\equalcont{These authors contributed equally to this work.}

\author[2]{\fnm{Alexander} \sur{Fuchs-Kreiß} \orcidlink{0000-0001-6483-8599}}\email{fuchskreiss@uni-hildesheim.de}
\equalcont{These authors contributed equally to this work.}

\affil*[1]{\orgdiv{School of Computer Science and Statistics}, \orgname{Trinity College Dublin}\orgaddress{\street{}, \city{Dublin}, \postcode{D02N8P5}, \state{Co. Dublin}, \country{Republic of Ireland}}}

\affil[2]{\orgdiv{Institute for Mathematics, Mathematics Education and Computer Science Education - IMMI}, \orgname{University of Hildesheim}, \orgaddress{\street{Samelsonplatz 1}, \city{Hildesheim}, \postcode{31141}, \state{Lower Saxony}, \country{Germany}}}





\maketitle
\begin{abstract}

Relational Event Models (REMs) provide a rigorous framework for analyzing dyadic interactions observed in continuous time, capturing history-dependent dynamics such as triadic closure and reciprocity.
Framing REMs through the lens of counting processes embeds the model in a rich theoretical foundation, facilitating its mathematical analysis. 
 While Maximum Likelihood Estimation (MLE) is standard practice for estimating these models, the underlying statistical guarantees rely on specific asymptotic regimes, namely, whether the network size ($n$), the observational period ($T$), or both approach infinity. 
We review the theoretical foundations of such counting-process-based models, formalizing the core assumptions required to achieve asymptotic normality across these different limits. 
With a specific focus on Cox-type multiplicative models, we detail the circumstances under which these assumptions hold. 
Supported by simulation studies, we illustrate how structural modeling choices, including temporal windowing and logarithmic transformations, affect empirical coverage and estimator convergence. 
We thereby derive several guiding principles for specifying such models in realistic contexts, bridging theory and practice. 
\end{abstract}


\section{Introduction}

With the advent of computational social science, continuously collected interaction data at scale, measuring, for instance,    
e-mail traffic via logged data or face-to-face interactions via sociometric badges, are becoming widely available \citep{lazer_computational_2009, wagner2021}. 
Contrary to survey data \citep{Bu08a}, we can easily measure such event data at scale and with fine-grained temporal resolution.
This data enables insights into micro behavior within organizations, such as hospitals \citep{kitts2017}, or in disrupted environments, such as the September 11 attacks on the World Trade Center \citep{renshaw2023}. 

The statistical unit in this context is a relational event, representing instantaneous interactions between entities at a given point in time.  
Denoting the set of observed entities by $\unitset \coloneqq \{1, \ldots, n\}$ and the observational period by $\timeset \coloneq [0, T]$ with $T>0$, a relational event is denoted by a tuple $e = (i,j,t)$ for an event that occurred between sender $i \in \unitset$ and receiver $j \neq i \in \unitset$ at time $t \in \timeset$. 
In the examples mentioned in the opening paragraph, an event $e = (i,j,t)$ corresponds to a patient transfer between hospitals $i$ and $j$ at time $t$ in \citet{kitts2017}, while it corresponds to communication between two responder groups during the World Trade Center attack in \citet{renshaw2023}.
We refer to the sender, receiver, and time of the $m$th event by $i_m, j_m,$ and $t_m$, respectively.
Events can more broadly be associated with a particular direction, duration, and weight \citep{fritz2025} and even occur between sets of entities \citep{espinosa_rada2025}. 
Most material developed in this manuscript generalizes to such settings, but for the sake of simplicity, we only cover undirected dyadic interactions in this article. 
The set of sender and receiver pairs $(i,j)$ is then given by $\dyadset \coloneqq \{(i,j)\mid i,j \in \unitset, i<j\}$.

We specify generative models for these dyadic events via $n \, (n-1)/2$-dimensional counting processes $\bm{N}(t) = (N_{i,j}(t))_{(i,j) \in \dyadset}$.
The $(i,j)$th coordinate of $\bm{N}(t)$ counts the number of observed events between entities $i$ and $j$ that were observed until and including any point in time $t \in \timeset$. 
To simplify later notation, we write $N_{i,j}(t) =N_{j,i}(t)$ for $i>j$. 
We characterize the instantaneous probability of a jump in any coordinate of these counting processes by the intensity  $\boldsymbol{\lambda}(t) = \left(\lambda_{i,j}(t)\right)_{(i,j) \in \dyadset}$.  
Different parametrizations for this intensity are surveyed in \citet{fritz2019} and \citet{bianchi2024}.

The common quantitative research pipeline starts by theorizing about potential statistics that inform interactions based on prior theoretical research, such as \citep{uzzi2004,uzzi1996}, that operationalize testable hypotheses. 
Neighboring entities in interaction networks are unlikely to behave independently, therefore, any realistic theory for such complex behavior  puts dependence between studied entities at its core \citep{WsFk94}.
For instance, if entities $i$ and $j$ have previously interacted with a common partner $h$, they might be more likely to interact with one another in the future \citep{hollandEvidenceTransitivityPositive1972}. 
In the words of \citet[p. 160]{Bu08a}, ``history creates the context for present action''.
An appropriate statistical model in this context can encode such dependencies in different model specifications. 
Both frequentist and Bayesian approaches can be used to draw inferential conclusions given the interaction data at hand. 
To our knowledge, most approaches within the Bayesian paradigm are methodological papers, e.g., \citep{fritz2021glitters, leifeld2026}, while the entirety of substantive application papers relied on frequentist approaches, e.g.,  \citep{kitts2017,renshaw2023, quintane2013, tranmer2015}.  
We therefore here focus on frequentist approaches where the parameters are estimated via Maximum Likelihood.

To carry out statistical tests to decide between different possible data-generating mechanisms or obtain confidence intervals, we heavily rely on the sampling distribution of the Maximum Likelihood estimate. 
Since this distribution is generally unavailable for finite samples, we need a central limit theorem to guarantee that the sampling distribution can be approximated by a multivariate normal distribution under some asymptotic regime.  
For event data studied in this article, one can think of different asymptotic regimes: 
\begin{enumerate}
    \item $T \rightarrow \infty$ with fixed $n$: Assume we are studying interactions among a fixed set of entities over some time and intend to draw inferential conclusions on where the estimates would lie if we were to observe the set for longer; 
    \item $n \rightarrow \infty$ with fixed $T$: Assume we are studying a growing population of entities over a fixed time and intend to draw inferential conclusions based on a subsample where all sufficient statistics only depend on other entities also in the sample.  
    \item $T$ and $n \rightarrow \infty$: Assume we are studying a growing population of entities over an increasing observational period. 
    This setting is the least explored for counting-based models and we need to assume some growth rates of $n$ and $T$. 
\end{enumerate}
While all asymptotic regimes can lead to a valid central limit theorem, different conditions must be met in each case. 
We argue that practitioners should carefully assess these while specifying their models and interpreting the results. 
We first describe  in Section \ref{sec:CPviewREM} how counting processes serve as a natural framework for modeling relational event data and then review in Section \ref{sec:practical_asymptotics} theoretical results.  
We here focus on the necessary assumptions and what they imply in practical settings. 
Using simulations and theoretical results, we illustrate in Section \ref{sec:simulation_study} how particular choices affect inference results, drawing on the survival analysis literature on independent observations.

\section{A Counting Process Framework for Modeling Relational Event Data}
\label{sec:CPviewREM}

Relational event data is a distinct type of time-to-event data in which events occur between pairs of entities, such as an email sent between them \citep{perry_point_2013}, rather than a single entity, such as a patient passing away. 
The most general framework for time-to-event data is based on counting processes developed by \citet{andersen1982}, which we adapt to relational event data.  
Introducing advanced mathematical concepts such as martingales, filtration, and predictability would make the review of more involved mathematical concepts necessary. Thus, we refer to \citet{ABGK93} and \citet{aalen2008} for a thorough treatment of these foundations. 

A counting process for pair $(i,j) \in \dyadset$ on the interval $\timeset$ is a stochastic process $N_{i,j}:\timeset \to\IN_0 \coloneqq\{0,1,2,...\}$, whose state at time $t \in \timeset$ is defined by 
\be
N_{i,j}(t) &=& \sum_{m \geq 1} \mathbb{I}(e_m = (i,j,t_m) \land t_m \leq t),
\label{eq:cp}
\ee
and counts the number of interactions between entities $i$ and $j$ with $(i,j) \in \dyadset$.
We distinguish between the entire stochastic process, $N_{i,j} \coloneqq \{N_{i,j}(t)\mid t\geq0\}$, and its status at $t \in \timeset$, $N_{i,j}(t)$, and collect the counting processes of all pairs of entities in $\bm{N} \coloneqq (N_{i,j})_{(i,j) \in \dyadset}$. 
By \eqref{eq:cp}, each coordinate of this multivariate counting process is nonnegative, integer-valued, and adapted to the natural filtration $\hist{t} \coloneqq \sigma(\{\bm{N}(u): u \leq t\})$, which is a so-called $\sigma$-field growing with time. 
One can think of $\hist{t}$ as the collection of all available information until and including time $t$. 
If additional information, such as covariate information for each unit or dyad, is available over time, we can augment the filtration $\hist{t}$ accordingly. 

Given these properties, one can show that $\bm{N}$ is a sub-martingale, allowing us to write its $(i,j)$th coordinate as a superposition of a  martingale and compensator process: 
\be
\label{eq:doob_meyer}
    N_{i,j}(t) &=&  M_{i,j}(t) + \Lambda_{i,j}(t), 
\ee
for all $t\in \timeset$ and $(i,j) \in \dyadset$. 
\eqref{eq:doob_meyer} is called the Doob-Meyer decomposition.
The martingale $M_{i,j}:\timeset\to\IR$ is a stochastic process such that $\IE(M_{i,j}(t) \mid \hist{s})=M_{i,j}(s)$ for all $s\leq t$. 
One can understand the martingale component as a residual or noise process without any structure related to the available information $\hist{t}$.
The compensator process, on the other hand, is the part of the counting process that is predictable with the available information. 

Since the compensator is a non-decreasing predictable process, we assume that it can be written
as an integral
\be 
\label{eq:cum_intensity}
\Lambda_{i,j}(t)=\dint_0^t\lambda_{i,j}(r)\dd r
\ee
for some non-negative stochastic process $\lambda_{i,j}:\timeset\to[0,\infty)$ for each $(i,j) \in \dyadset$, called intensity.
Thus, we here implicitly assume that the process defined through \eqref{eq:cum_intensity} is absolutely continuous. 
In case the compensator $\Lambda_{i,j}$ experienced a jump at $t_0$ when $N_{i,j}(t_0)$ experiences a jump with positive probability at $t_0$, it would prohibit the integral representation in \eqref{eq:cum_intensity}. 
Therefore, we exclude such discrete behavior and focus on the continuous case.
The matrix $\bm{\lambda} = (\lambda_{i,j})_{(i,j) \in \dyadset}$ defines in its $(i,j)$th coordinate the instantaneous probability of a jump in the counting process $N_{i,j}(t)$: 
\be
\label{eq:intensity2}
 \lambda_{i,j}(t) = \lim_{h \downarrow 0} \frac{1}{h} \mathbb{P} \left( N_{i,j}(t+h) - N_{i,j}(t) = 1 \mid \hist \right), 
\ee
where $t^-$ denotes the time point right before $t \in \timeset$. 
The information from $t$ is here excluded since what happens at instance $t$ should not affect $\lambda_{i,j}(t)$. 

While the formulation in \eqref{eq:intensity2} is nonparametric and potentially dependent on the complete filtration $\hist$, in most real-life applications it is more practical to assume that some low-dimensional vector encodes all necessary information from the past. 
To describe this low-dimensional information, we introduce the sufficient statistics $\covs_{i,j}: \timeset \to \mathbb{R}^p$ to map the high-dimensional historical information to a $p$-dimensional covariate space. 
We omit the specific dependence on $\hist$ and write $\covs_{i,j}(t) = (s_{i,j, 1}(t), ..., s_{i,j, p}(t))^\top$ but keep in mind that $\covs_{i,j}(t)$ may only depend on information that is observable before time $t$. 
These statistics summarize all dependence of $\lambda_{i,j}(t)$ on the past and facilitate the integration of substantive theory in model specification. 
For example, the theory of triadic clustering \citep{hollandEvidenceTransitivityPositive1972} can be assessed by incorporating the count of common partners of pair $(i,j) \in \dyadset$ in the $q$th coordinate of the sufficient statistics via $s_{i,j, q}(t) = \sum_{h \notin \{i,j\}} \mathbb{I}(N_{i,h}(\timenow) > 0 )\, \mathbb{I}(N_{h,j}(\timenow) >0)$ for some $q \in \{1, ... p\}$. 
Similarly, the conversational norm of reciprocal communication  \citep{gibson2005} can be captured in the statistic $s_{i,j, q}(t)  = \mathbb{I}(\Tilde{e}_t = (j, i))$, with $\Tilde{e}_t$ denoting the tuple of sender and receiver of the last event observed before time $t$.
Consequently, sufficient statistics $\covs_{i,j}(t)$ are the primary tool for encoding arbitrary dependence assumptions of past events on future events.   



The final model component is the specification of the structural relationship between $\covs_{i,j}(t)$ and $\lambda_{i,j}(t)$. 
First, 
we transform the sufficient statistics $\covs_{i,j}(t)$ to a scalar value $\eta_{i,j,t} = \bm{\theta}^\top \covs_{i,j}(t)$, where $\bm{\theta} = (\theta_1, ..., \theta_p) \in \mathbb{R}^p$ is a $p$-dimensional unknown parameter vector. 
Second, we transform this value by a function $g: \mathbb{R} \rightarrow \mathbb{R}^+$ that is assumed to be strictly positive, bounded away from zero, and three times continuously differentiable. 
The intensity is then given by: 
\be
\label{eq:intensity_rem}
      \lambda_{i,j}\left(t \mid \bm{\theta} \right) &=& g\left(\eta_{i,j,t}\right),
\ee
which generalizes several prominent model classes within the literature. 
In this paper, we assume a correctly specified model, such that there exists a true parameter $\truepara$ such that $(\lambda_{i,j}(t\mid\truepara))_{(i,j) \in \dyadset}$ is equal to the intensity functions associated with $\bm{N}$.

\begin{example}[Additive Models]
\label{ex:additive}
Letting $g(x) = x$ and defining the sufficient statistic vector as $\covs_{i,j}(t) = (1, \int_0^t \exp(-(t-u)) \dd N_{i,j}(u))^\top$ yields a Hawkes process for the interaction data \citep{hawkes1971}. 
In this example, only previous events between entities $(i,j) \in \dyadset$ affect the intensity $\lambda_{i,j}\left(t \mid \bm{\theta} \right)$ additively in the near future, where the past is exponentially down-weighted. 
We need to make sure that $\covs_{i,j}(t)^\top \theta > 0$ holds for all $(i,j) \in \dyadset$ and $t \in \timeset$. 
Several comparable methods were proposed in the literature in this additive setting \citep{passino2023, vuContinuousTimeRegressionModels2011, soliman2022a,fangGroupNetworkHawkes2024}. 
\end{example}

\begin{example}[Multiplicative Models]
\label{ex:multiplicative}
Setting $g: x\mapsto \exp(x)$, yields the original Relational Event Models as discussed in \citet{butts2008}.  
The effect of the $q$-th coordinate $s_{i,j,q}(t)$ of the sufficient statistics increasing by one unit, is then given by the multiplicative factor $\exp(\theta_q)$. 
Much literature on such Cox-type models has been developed \citep{KMP19, vuDynamicEgocentricModels2011, Fritz2020a, fritz2021glitters}. 
\end{example}

\section{Practical Asymptotics}
\label{sec:practical_asymptotics}

The choice of sufficient statistics reflects competing hypotheses regarding the underlying data-generating process. 
In applied research, selecting the most appropriate specification requires a principled statistical framework. 
Standard inferential tools, such as confidence intervals and hypothesis tests, typically rely on the asymptotic distribution of the estimator $\hat{\bm{\theta}}$. 
In this section, we will discuss, how the asymptotic regimes where either $T\to\infty$ or $n\to\infty$ or both, affect the mathematical analysis. 
To keep our notation general, we introduce the sample size $\sampsize:=\sampsize(n,T)$ and write $\sampsize\to\infty$ for either $n\to\infty$, $T\to\infty$ or both. 
In the classical setting of iid data, we would expect that $T$ is fixed and $n\to\infty$. 
However, in an REM context, it is not always reasonable to assume that units behave independently. 
But, as we will point out, this is not necessarily a problem as long as behavior of a new unit brings new information that can be leveraged for estimating the true data-generating parameter denoted by $\truepara$. 
On the other hand, from a time-series point of view, we would expect that $n$ is fixed and $T\to\infty$. In such a setting, observing the units for longer times provides more information for estimating $\truepara$. 
In this context, dependence between units is natural, but a certain stationarity in time will be required. 

Our focus lies on the maximum likelihood estimator
\be
\label{eq:mlest}
\MLE:=\underset{\para\in\paraspace}{\textrm{argmax}}\,\ell_{\sampsize}(\para), 
\ee
where the log-likelihood of the counting-process based models developed above is 
\beno 
\ell_\sampsize(\para):=\dsum_{_{(i,j) \in \dyadset}}\left(\dint_\timeset \log\lambda_{i,j}(t\mid\para)\dd N_{i,j}(t)-\int_\timeset \lambda_{i,j}(t\mid\para)\dd t\right). 
\ee
In practice, the maximizer in \eqref{eq:mlest} can often be found via iterative methods, such as Newton-Raphson \citep{vuDynamicEgocentricModels2011}, Fisher scoring \citep{butts2008}, or Expectation Maximization algorithms \citep{passino2023}. 
In Section~\ref{subsec:general_asymptotics}, we will start to discuss the general mathematical starting point to obtain an asymptotic distribution of the resulting Maximum Likelihood estimator, independent of the exact regime of interest. 
In Section~\ref{subsec:Cox}, we then emphasize for the case of a multiplicative Cox-type model, how the regimes affect the mathematical analysis.

 To explain the relevance of the assumptions that will be mentioned below, we will use the following definitions 
\begin{align}
 U_\sampsize(\para)\coloneqq &\partial_{\para}\ell_\sampsize(\para)=\sum_{(i,j) \in \dyadset}\left(\int_\timeset \frac{\partial_{\para}\lambda_{i,j}(t\mid\para)}{\lambda_{i,j}(t\mid\para)}\dd N_{i,j}(t)-\int_\timeset \partial_{\para}\lambda_{i,j}(t\mid\para)\dd t\right), \label{eq:score} \\
H_\sampsize(\para)\coloneqq&-\partial_{\para}^2\ell_\sampsize(\para), \nonumber \\
F_\sampsize(\para)\coloneqq&\sum_{(i,j) \in \dyadset}\int_\timeset \left(\frac{\partial_{\para}\lambda_{i,j}(t\mid\para)}{\lambda_{i,j}(t\mid\para)}\right)\left(\frac{\partial_{\para}\lambda_{i,j}(t\mid\para)}{\lambda_{i,j}(t\mid\para)}\right)^\top\lambda_{i,j}(t\mid\para)\dd t, \label{eq:F}
\end{align}
where $\partial_{\para}$ and $\partial_{\para}^2$ denote the gradient and Hessian matrix with respect to $\para$, respectively\footnote{Hence, we implicitly assume that $g$ is sufficiently often differentiable.}. 
Note that $\log\lambda_{i,j}(t\mid\para)$  appears as the derivative $\partial_{\para} \log \lambda_{i,j} = \partial_{\para} \lambda_{i,j}/\lambda_{i,j}$ in integrals with respect to $N_{i,j}$ or when multiplied with $\lambda_{i,j}(t\mid\truepara)$. 
Thus, we restrict all integrals to $\{t\in[0,T]:\,\lambda_{i,j}(t\mid\truepara)>0\}$ to avoid taking the logarithm of $0$.
Typically, $U_\sampsize$ and $H_\sampsize$ are called score function and observed information, respectively. The random variable $F_\sampsize$ is related to the variance of the score function and we will discuss later in which cases both $H_\sampsize$ and $F_\sampsize$ converge to the Fisher information.

\subsection{Asymptotic normality for parametric models}
\label{subsec:general_asymptotics}
We give now relatively straightforward assumptions that one has to check to obtain asymptotic normality of the maximum likelihood estimator $\hat{\para}_\sampsize$ in a parametric REM. 

\begin{assumptionA}[Smoothness and bounded statistics]
\label{assump:bounded_stats}
The function $g: \mathbb{R} \rightarrow(0,\infty)$ is three times continuously differentiable. The parameter space $\paraspace\subseteq\IR^p$ is bounded. Moreover, the sufficient statistics are uniformly bounded
, that is,
$$\sup_{n\in\IN}\sup_{t\in\timeset}\max_{(i,j) \in \dyadset}\|\covs_{i,j}(t)\|\leq M\text{ almost surely}$$
for some $M<\infty$.
\end{assumptionA}
\hide{
\alertA{Oben stehe einige Dinge, die schon in Kapitel 2 auftauchen, aber ich finde es gut, wenn alle Annahmen auch als solche gekennzeichnet sind. Deshalb habe ich das hier nochmal wiederholt.}
}

\begin{assumptionA}[Convergence of observed information]
\label{assump:Fisher}
For $a_\sampsize=\sqrt{Tn(n-1)/2} \rightarrow \infty$ it holds that
\beno
a_\sampsize^{-2}\dsum_{(i,j) \in \dyadset}\dint_\timeset \left(\partial_{\para_k}\partial_{\para_l}\log\lambda_{i,j}(t\mid\truepara)\right)^2\lambda_{i,j}(t;\truepara)\dd t
\ee
converges in probability to a finite constant for all $k,l\in\{1,...,p\}$. There is a positive definite matrix $\Sigma(\truepara)\in\IR^{p\times p}$ such that
$$a_\sampsize^{-2}F_\sampsize(\truepara)\overset{\IP}{\to}\Sigma(\truepara).$$
\end{assumptionA}

\begin{assumptionA}[Scaling]
\label{assump:scaling}
The sequence
\beno 
a_\sampsize^{-2}\dsum_{(i,j) \in \dyadset}\dint_\timeset \lambda_{i,j}(t\mid\truepara)\dd t
\ee
converges in probability to a finite constant.
\end{assumptionA}

The smoothness requirements on $g$ in Assumption~\ref{assump:bounded_stats} are often fulfilled in practice and ensure that some technical conditions are met. 
The boundedness of the sufficient statistics can, for example, be achieved by rescaling. We return to this point in remark \ref{rm:scaling}. 
The choice of $a_\sampsize$ in Assumption~\ref{assump:Fisher} is natural in view of the required convergences because $n(n-1)/2$ equals the number of unordered pairs and $T$ is the length of the integral. It will turn out in Theorem~\ref{thm:AN} below that $a_\sampsize$ determines the rate of convergence of $\MLE$ to $\truepara$. 
The limit $\Sigma(\truepara)$ in Assumption~\ref{assump:Fisher} plays the role of the Fisher information, and the two convergences in Assumption~\ref{assump:Fisher} together imply that
\begin{equation}
    \label{eq:Hesseconv}
    a_\sampsize^{-2}H_\sampsize(\truepara)\overset{\IP}{\to}\Sigma(\truepara)
\end{equation}
for $\sampsize\to\infty$, which we would expect because we consider correctly specified models. 
Assumption~\ref{assump:Fisher} is the main requirement that one has to check for a specific model. We provide more details in case of the Cox model in Section~\ref{subsec:Cox}. 
The final Assumption~\ref{assump:scaling} is posed for technical reasons. 
Given Assumption~\ref{assump:Fisher}, it does not pose any substantial constraint because the convergence requirements of Assumption~\ref{assump:Fisher} are very similar to those of Assumption~\ref{assump:scaling}. 
For the Cox-type model discussed in Section \ref{subsec:Cox}, we even observe that Assumption~\ref{assump:scaling} is not required.

Under these assumptions, the following theorem guarantees asymptotic normality, as a direct consequence of the general results in \citet{ABGK93}.

\begin{theorem}[Theorems VI.1.1 and VI.1.2 \cite{ABGK93}]
\label{thm:AN}
Let Assumptions~\ref{assump:bounded_stats}-\ref{assump:scaling} hold. Then, the maximum likelihood estimator $\MLE$ is consistent and
$$a_\sampsize\left(\MLE-\truepara\right)\overset{d}{\to}\mathcal{N}(0,\Sigma(\truepara)^{-1}).$$
\end{theorem}

We discuss in Appendix~\ref{app:discussion_HLassumptions} how our Assumptions~\ref{assump:bounded_stats}-\ref{assump:scaling} imply the conditions in \citet{ABGK93}. 
To give some insight into the proof, we provide a short discussion here. Assumptions~\ref{assump:bounded_stats} and~\ref{assump:Fisher} imply that $U_\sampsize(\truepara)$ behaves asymptotically like a normal distribution. 
Plugging the Doob-Meyer decomposition from \eqref{eq:doob_meyer} into \eqref{eq:score}, we obtain
\begin{align*}
U_\sampsize(\truepara)=&\sum_{(i,j) \in \dyadset}\Bigg(\int_\timeset \frac{\partial_{\para}\lambda_{i,j}(r\mid\truepara)}{\lambda_{i,j}(r\mid\truepara)}\dd M_{i,j}(r)+\int_\timeset \frac{\partial_{\para}\lambda_{i,j}(r\mid\truepara)}{\lambda_{i,j}(r\mid\truepara)}\lambda_{i,j}(r\mid\truepara)\dd r \\
&\qquad-\int_\timeset \partial_{\para}\lambda_{i,j}(r\mid\truepara)\dd r\Bigg)=\sum_{(i,j) \in \dyadset}\int_\timeset \frac{\partial_{\para}\lambda_{i,j}(r\mid\truepara)}{\lambda_{i,j}(r\mid\truepara)}\dd M_{i,j}(r).
\end{align*}
The last stochastic integral is a martingale because its integrator is a martingale itself.
Therefore, Rebolledo's martingale central limit theorem (e.g., Theorem~II.5.1 in \cite{ABGK93}) implies under Assumptions~\ref{assump:bounded_stats} and~\ref{assump:Fisher} that $a_\sampsize^{-1}U_\sampsize(\truepara)$ converges to a $p$-dimensional normal distribution $\mathcal{N}(0,\Sigma(\truepara))$.  To transfer this asymptotic normality to the estimator $\hat{\theta}_{\sampsize}$, one essentially applies a Taylor expansion and proves that the remainder term converges to zero as $\sampsize \rightarrow \infty$. More details on this topic are provided in Appendix~\ref{subsec:proof_idea}.

The statement of Theorem~\ref{thm:AN} holds for all asymptotic regimes; however, the growth rate of $a_\sampsize$ and the validity of the assumptions depend on it. 
The main question is if Assumption~\ref{assump:Fisher} holds. 
In principle, it can hold either if $n\to\infty$, $T\to\infty$, or both. 
Thus, estimation of the parameter $\truepara$ is possible in large networks that are observed over a short time period, or in small networks that are observed for a long time. 
From a substantive perspective, which asymptotic framework is more reasonable depends on the data at hand and what type of inferential conclusions should be drawn from it, see \citet{mccullagh2002} for a wider discussion. 
From a mathematical perspective, Assumption~\ref{assump:Fisher} can hold for $T\to\infty$ only if the averages over time of the intensities stabilize. 
If one wants to employ an $T\to\infty$ assumption, one has to believe that the behavior in different (far apart) time intervals is only weakly correlated. 
Conversely, in the case $n\to\infty$, spatial averages are required to stabilize, i.e., one has to be comfortable assuming that different (far apart) sub-networks are only weakly correlated.

\begin{remark}[Feasible Set of Interactions]
We assumed here that interactions are possible between all pairs $(i,j) \in \dyadset$.
One may exclude certain pairs from the interactions by studying intensities of the form $C_{i,j}(t)\, g(\para^\top\covs_{i,j}(t))$, where $C_{i,j}(t)\in\{0,1\}$ indicates if the pair $(i,j)$ is at risk of an event at time $t$.
The discussion in this section, can be extended to this case by choosing a different sequence $a_\sampsize$.
For details we refer to the discussion in \citet{KMP19}.
\end{remark}

\begin{remark}[Scaling of Sufficient Statistics]
\label{rm:scaling}
When dealing with sufficient statistics that are not uniformly bounded (e.g., those growing with $\log n$ or even $n$), one may standardize the covariates prior to estimation to satisfy Assumption~\ref{assump:bounded_stats}. 
However, scaling the sufficient statistics by a sequence $d_\sampsize \to \infty$ requires the true parameter $\truepara$ to change with $\sampsize$ to absorb this scaling. 
Therefore, the unscaled coefficient must shrink to zero at the rate $\mathcal{O}(d_\sampsize^{-1})$ as the network grows and $d_\sampsize\times a_\sampsize$ determines the convergence rate of $\truepara$. 
As an alternative, one may assume the true parameter $\truepara$ to remain fixed. In that case, the change in scale in $\covs_{i,j}(t)$ can instead be controlled by standardizing the score and the observed information. 
We provide more details how this works in Appendix~\ref{subsubsec:scaling} and give only a brief discussion here.
Assume for simplicity that
\beno 
v_{\sampsize,k} &\coloneqq &\IE\left(\left(\partial_{\para_k}\lambda_{i,j}(t\mid\truepara)\right)^2\lambda_{i,j}(t\mid\truepara)\right)
\ee
is independent of $t$ for $k=1,...,p$.
Let $\mathbf{D}_\sampsize$ be a diagonal matrix with diagonal entries $(\sqrt{v_{\sampsize,k}})_{k=1}^p$.
One can use this matrix to scale the score vector and the observed information (e.g., $\mathbf{D}_\sampsize^{-1} H_\sampsize(\truepara) \mathbf{D}_\sampsize^{-1}\overset{\IP}{\to}\Sigma(\truepara)$) after the derivatives have been computed. This preserves the Central Limit Theorem in the sense that\footnote{We give more details about the necessary assumptions in Appendix~\ref{subsubsec:scaling}. But we note here that in particular Lemma~\ref{lem:ABGK:bound} is potentially no longer applicable to prove Assumptions~\ref{assump:ABGK:Hesse_uniform} and~\ref{assump:ABGK:consistency} in Appendix~\ref{app:discussion_HLassumptions} and other tools have to be used instead.}
\beno 
a_\sampsize\mathbf{D}_\sampsize\left(\MLE-\truepara\right)&\overset{d}{\to} & \mathcal{N}\left(0,\Sigma(\truepara)^{-1}\right).
\ee
Note, however, that assuming that the unscaled coefficient stays fixed can imply explosive behavior of the counting processes. 
Simulating from such a model can imply so-called exploding counting processes. 
A non-explosive counting process must satisfy the Feller criterion for any finite sample size $\mathcal{S}$:
\be
\label{eq:feller}
\dsum_{k =1}^\infty \left(\dsum_{(i,j) \in \dyadset} g(\bm{\theta}^\top \covs_{i,j}(t_k))\right)^{-1} = \infty. 
\ee
Intuitively, one can comprehend \eqref{eq:feller} as guaranteeing that the waiting time until the $\infty$th event is $\infty$. 
For $\covs_{i,j}(t) \in \mathbb{R}^p$, it is, in practice, easier to define an envelope  $g(\bm{\theta}^\top \tilde{\covs}(T,n)) \geq g(\bm{\theta}^\top \covs_{i,j}(t_k))$ for all $(i,j) \in \dyadset$ and $t \in \timeset$ and substitute \eqref{eq:feller} with 
\beno
\dsum_{k =1}^\infty \left(\dsum_{(i,j) \in \dyadset} g(\bm{\theta}^\top \tilde{\covs}(T,n))\right)^{-1} = \infty,
\ee
which is called the Jacobsen condition. 
Else, the Doob-Meyer decomposition from \eqref{eq:doob_meyer} and the foundation of this mathematical framework is invalid \citep{gjessing2010}. 
\end{remark}

\subsection{Application to Cox-type Models}
\label{subsec:Cox}

In this section, we discuss the multiplicative case mentioned in Example \ref{ex:multiplicative} with $g(x):=\exp(x)$ and $\lambda_{i,j}(t\mid\para)=\exp(\para^\top\covs_{i,j}(t))$.  
To fulfill Assumption~\ref{assump:bounded_stats}, we assume that $\paraspace$ is bounded and that $\covs_{i,j}(t)$ is uniformly bounded. 
The differentiability of $g$ is guaranteed by the choice $g(x)=\exp(x)$. For the first part of Assumption~\ref{assump:Fisher}, note that $\log\lambda_{i,j}(t\mid\para)=\covs_{i,j}(t)^\top\para$ with
$$\partial_{\para_k}\log\lambda_{i,j}(r\mid\truepara)=\covs_{i,j}^{(k)}(t).$$
This implies that $\partial_{\theta_k}\partial_{\theta_l}\log\lambda_{i,j}(t\mid\para)=0$ and that the first part of Assumption~\ref{assump:Fisher} holds. 
The convergence in Assumption~\ref{assump:scaling} is actually not required in this case: 
Upon closer inspection of Appendix~\ref{app:discussion_HLassumptions} one can observe that Assumption~\ref{assump:scaling} is only required for proving that the condition on $H_{i,j}$\footnote{$H_{i,j}$ bounds $\partial_{\para_k}\partial_{\para_l}\partial_{\para_m}\log\lambda_{i,j}(t\mid\para)=0$.} from Lemma~\ref{lem:ABGK:bound} holds. 
However, in the Cox case, $H_{i,j}\equiv0$, that is, we can directly check the condition on $H_{i,j}$ from Lemma~\ref{lem:ABGK:bound} without further assumptions. 
Note that the above discussion is valid for all asymptotic regimes.

We are left to discuss the second part of Assumption~\ref{assump:Fisher}, that is, the convergence of $a_\sampsize^{-2}F_\sampsize(\truepara)\to\Sigma(\truepara)$ for a positive definite matrix $\Sigma(\truepara)$. We rewrite
\begin{align*}
    F_\sampsize(\truepara)=&\sum_{(i,j) \in \dyadset}\int_\timeset \covs_{i,j}(t)^\top\covs_{i,j}(t)\exp(\covs_{i,j}(t)^\top\truepara)\dd t= \sum_{(i,j) \in \dyadset}\int_\timeset  F_{i,j}(t)\dd t,
\end{align*}
where  $F_{i,j}(t):=\covs_{i,j}(t)^\top\covs_{i,j}(t)\exp(\covs_{i,j}(t)^\top\truepara)$. 
Consider first the case where $n$ is fixed and $T\to\infty$. 
In this case, $\overline{F}_n(t):=2/(n(n-1))\sum_{(i,j) \in \dyadset}F_{i,j}(t)$ can be understood as a single stochastic process evolving over $t \in \timeset$. 
We require convergence of $1/T\int_\timeset \overline{F}_n(t)\dd t$, which follows if the process $\overline{F}_n$ has certain ergodic properties\footnote{See Chapter 25 of \cite{K21} for an in-depth treatment.}. 
Such a property is, for instance, implied by a fading memory in the sense that characteristics of $\overline{F}_n$ on a certain time interval $I_1 \subset \timeset$ behave almost independently of characteristics of $\overline{F}_n$ on another time interval $I_2\subset \timeset$ when $I_1$ and $I_2$ drift further apart.
In applied research, two ways are common implemented to guarantee such a fading memory: windowed statistics are employed in \citet{stadtfeld2017} where events can only affect the intensity for a specific amount of time, whereas \citet{Brandes2009} assumed that the effect of interactions at time $s$ for the intensity at time $t>s$ is given by the exponential function with a particular half time. 

For fixed $n$, the dependence structure of $(F_{i,j})_{(i,j)\in \dyadset}$ is not irrelevant but plays a less prominent role.
On the other hand, if $n\to\infty$, justifying ergodicity purely from the temporal behavior becomes more difficult as the summation over $(i,j) \in \dyadset$ in \eqref{eq:F} can introduce an additional drift. 
In fact, if $T$ is finite, such an argument becomes even impossible, and one has to rely on a certain ergodicity property in location.  
Writing
\beno 
a_\sampsize^{-2}F_\sampsize(\truepara)&=&\dfrac{2}{n(n-1)}\dsum_{(i,j)\in\dyadset}\tilde{F}_{i,j}
\ee
with $\tilde{F}_{i,j}:=\frac{1}{T}\int_0^TF_{i,j}(t)\dd t$,
we now require the average over dyads, $\sum_{(i,j)\in\dyadset}$, instead of the integral, $\int_\timeset$, to stabilize.
We illustrate how such a property can be derived from a weak correlation property for disjoint dyads in the case $n\to\infty$ and $T$ fixed, using ideas from \cite{KMP19}, and then motivate weak correlation by introducing a fading memory in location. 

Suppose that the whole process is vertex exchangeable, that is, the joint distributions of $(N_{i,j},\covs_{i,j})_ {(i,j) \in \dyadset}$ and $(N_{\pi(i),\pi(j)},\covs_{\pi(i)\pi(j)})_{(i,j) \in \dyadset}$ are the same for any permutation $\pi$ of $\{1,...,n\}$.
In particular, this assumption implies that $(N_{i,j},\covs_{i,j})_{(i,j)\in\dyadset}$ is a collection of identically distributed random variables.
A simple, and in this case sufficient, strategy to prove convergence of $a_\sampsize^{-2}F_\sampsize(\truepara)$ relies on the Markov inequality to check the definition of convergence in probability.
For illustration, we assume that $\covs_{i,j}(t)\in\IR$, but, on the expense of more involved notation, the argument transfers to the multivariate case. For arbitrary $\epsilon>0$, it holds that
\begin{align}
&\IP\left(\left|a_\sampsize^{-2}F_\sampsize(\truepara)-\frac{1}{T}\int_\timeset \IE(F_{1,2}(t))\dd t\right|>\epsilon\right) \nonumber \\
\leq\,&\,\frac{1}{\epsilon^2}\,\IE\left(\left|a_\sampsize^{-2}F_\sampsize(\truepara)-\frac{1}{T}\int_\timeset \IE(F_{1,2}(t))\dd t\right|^2\right) \nonumber \\
=&\,\frac{1}{\epsilon^2}\IE\left(\left|a_\sampsize^{-2}\sum_{(i,j) \in \dyadset}\int_\timeset \left(F_{i,j}(t)-\IE(F_{i,j}(t))\right)\dd t\right|^2\right) \nonumber \\
=&\,\frac{1}{\epsilon^2a_\sampsize^4}\sum_{(i,j) \in \dyadset}\textrm{Var}\left(\int_\timeset F_{i,j}(t)\dd t\right) \nonumber \\ 
&\quad+\frac{1}{\epsilon^2a_\sampsize^4}\sum_{(i,j)\in\dyadset}\sum_{(k,l)\in\dyadset\setminus\{(i,j)\}}\textrm{Cov}\left(\int_\timeset F_{i,j}(t)\dd t,\int_\timeset F_{k,l}(t)\dd t\right) \nonumber \\
=&\,\frac{2}{\epsilon^2\cdot n(n-1)}\textrm{Var}\left(\frac{1}{T}\int_\timeset F_{1,2}(t)\dd t\right) \label{eq:two_overlap} \\
&\quad+\frac{2(n-2)}{\epsilon^2\cdot n(n-1)}\textrm{Cov}\left(\frac{1}{T}\int_\timeset F_{1,2}(t)\dd t,\frac{1}{T}\int_\timeset F_{1,3}(t)\dd t\right) \label{eq:one_overlap} \\
&\quad+\frac{(n-2)(n-3)}{\epsilon^2\cdot n(n-1)}\textrm{Cov}\left(\frac{1}{T}\int_\timeset F_{1,2}(t)\dd t,\frac{1}{T}\int_\timeset F_{3,4}(t)\dd t\right), \label{eq:no_overlap}
\end{align}
where the last equality follows because, by vertex exchangeability, the covariance terms depend only on the amount of overlap between the pairs $(i,j)$ and $(k,l)$\footnote{There are $n(n-1)/2$ possibilities to chose $(i,j)$, $n(n-1)(n-2)/2$ possibilities to choose $(i,j)$ and $(k,l)$ with one overlapping vertex, and $n(n-1)(n-2)(n-3)/4$ possibilities to choose $(i,j)$ and $(k,l)$ without overlap.}.

If we can show that Equations \eqref[s]{eq:two_overlap}-\eqref[s]{eq:no_overlap} converge to $0$ for all $\epsilon>0$ when $n\to\infty$, and that $\Sigma(\truepara):=1/T\int_\timeset \IE(F_{1,2}(t))\dd t$ is invertible\footnote{Recall that $T$ is fixed. 
However, if $T\to\infty$, assuming the convergence of this expression is the only extra assumption. 
All of the following arguments will remain true. Note furthermore that $\Sigma(\truepara)$ is always positive semi-definite.}, we obtain Assumption~\ref{assump:Fisher}. 
Recall that $F_{i,j}(t)$ is a function of $\covs_{i,j}(t)$, thus, such requirements are effectively concerning the statistics $(\covs_{i,j})_{(i,j)\in\dyadset}$.
We see that Equations \eqref[s]{eq:two_overlap} and \eqref[s]{eq:one_overlap} converge to zero under mild assumptions because the factor in front of the variance (covariance) converges to zero. We provide a detailed discussion of these assumptions in Appendix~\ref{app:Cox}.

In \eqref{eq:no_overlap}, the factor in front of the covariance converges to a non-zero constant, i.e., we require that the covariance over disjoint dyads converges to zero. 
To give a heuristic motivation for such an assumption, suppose that there is an underlying, potentially unknown, sparse network $G$ that contains all dyads $(i,j)\in\dyadset$ as vertices. We assume that the amount of dependence between $\covs_{i,j}$ and $\covs_{k,l}$, and hence $F_{i,j}$ and $F_{k,l}$, decays as the shortest path length distance in $G$, denoted by $d((i,j),(k,l))$, increases. 
The amount of dependence could, e.g., be measured in terms of correlation of $F_{i,j}(t)$ and $F_{k,l}(t)$. 
In such scenarios, it is plausible that
\beno 
\IE(F_{1,2}(t)F_{3,4}(s)\mid d((1,2),(3,4))=k) &\approx&\IE(F_{1,2}(t))\IE(F_{3,4}(s)),
\ee
when $k$ is large.
Since, in sparse networks, it is much more likely that $d((1,2),(3,4))$ is large, these cases dominate, and we may expect the covariance in \eqref{eq:no_overlap} to converge to zero. We give more details in Appendix~\ref{app:Cox}. Thus, we may expect that $a_\sampsize F_\sampsize(\truepara)$ converges and Assumption~\ref{assump:Fisher} can be expected to hold. 

In summary, we have shown that Assumptions \ref{assump:bounded_stats}-\ref{assump:scaling} are satisfied as long as the sufficient statistics
are defined with caution: 
For $T \rightarrow \infty$, they should have a fading memory with passing time, for $n \rightarrow \infty$,
they should not induce global dependencies.

\section{Simulation Study}
\label{sec:simulation_study}

We assess the finite-sample validity of the theoretical results from Section \ref{sec:practical_asymptotics} through Monte Carlo simulations. 
The $\mathtt{R}$ package $\mathtt{redeem}$ facilitates both data generation and model estimation \citep{redeem}. 
We adopt the multiplicative Cox-type specification detailed in Section \ref{subsec:Cox}, where the vector of sufficient statistics comprises an intercept alongside selected endogenous sufficient statistics.
To test the boundaries of our asymptotic framework, the study includes model specifications that intentionally violate the stability conditions derived in Section \ref{subsec:Cox}.
To model varying degrees of temporal dependence, we construct both windowed and cumulative formulations of these statistics. 
We apply identity and $\log(\cdot + 1)$ transformations to these statistics to evaluate the sensitivity of estimator stability.
 This yields four distinct model specifications per simulation setting: cumulative versus windowed statistics (orange and grey in subsequent visualizations) with identity versus log transformations (bright and dark).
 Because the scales across those statistics are not directly comparable, we restrict the true parameter values to the positive real line, $\mathbb{R}_{+}$, and provide their exact numerical specifications in Appendix \ref{sec:sim_setup}. 

\begin{figure}[t!]
        \centering
        \includegraphics[width=0.75\linewidth]{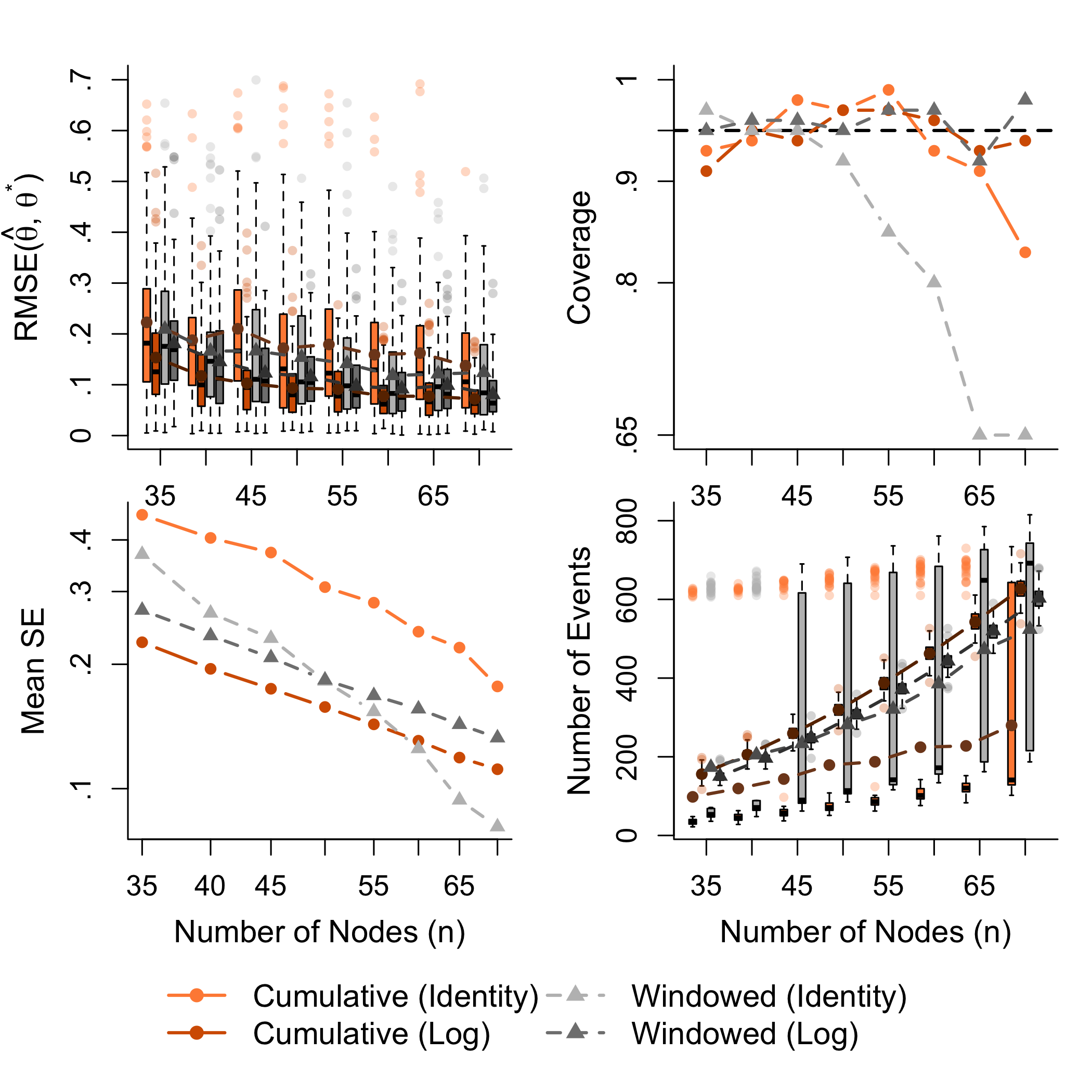}
        \caption{Simulation Study with $n\rightarrow \infty$: Left panels display the estimator RMSE and Mean Standard Error; right panels display empirical coverage probabilities and event counts.}    
                \label{fig:n}
\end{figure}

The Root Mean Square Error (RMSE) measures consistency of the point estimates. 
The mean standard error serves as a proxy for how fast the model accumulates information, while empirical coverage probabilities validate the variance estimator and confirm the adequacy of the Gaussian approximation in finite samples. 
Finally, we monitor the total number of events to detect explosive behavior in the counting processes. 

\hide{
At the same time, we set $f$ to be the identity, $f: x \mapsto x$, and logarithm, $f: x \mapsto \log (x+1)$, to examine how different scaling of the covariates, which was mentioned in Remark \ref{rm:scaling}, affects the simulations. 
The true parameter $ = (\theta_{\text{intercept}}, \theta_{\text{SP}})^\top$ is set to .. . 
$\covs_{i,j}(t) = (1, f(\sum_{k \notin \{i,j\}} A_{i,k}(t, \Delta) A_{k,j}(t, \Delta)))^\top$ for a set function $f$ with $A_{i,j}(t, \Delta)$ denoting the indicator function whether an event for pair $(i,j) \in \dyadset$ was observed in the interval $[t- \Delta, t)$. 
We vary $\Delta$ to cover settings where the temporal dependence between covariates is weak ($\Delta = 0.5$) or strong ($\Delta = T$).
}

\noindent\textbf{Scenario: $n \rightarrow \infty$:}
We employ the inertia statistic as the endogenous statistic, counting the number of prior interactions for each pair. 
In Figure \ref{fig:n}, we evaluate the asymptotic behavior of the estimator as the cardinality of $\unitset$ expands from 35 to 70  entities for a fixed $T$. 
As $|\unitset|$ increases, both the Root Mean Square Error (RMSE) and the theoretical standard errors decrease for all specifications.
In line with Remark \ref{rm:scaling},
the model involving the raw count of prior interactions  accumulates information faster than the log-transformed specification. 
Approximately $30\%$ of these simulations display explosive counting processes
\footnote{It is straightforward to see that here, \eqref{eq:feller} is not satisfied. In our simulations, we define a counting process to be exploding whenever any dyadic intensity exceeds $1.79769 \times 10^{308}$, indicating overflow.}, leading to improper empirical coverage and extreme variance in the total event count across simulation runs. 
The $\log(\cdot + 1)$ transformation yields well-behaved, stable stochastic processes,  highlighting the need for transforming raw counts.

\begin{figure}[t!]
        \centering
        \includegraphics[width=0.75\linewidth]{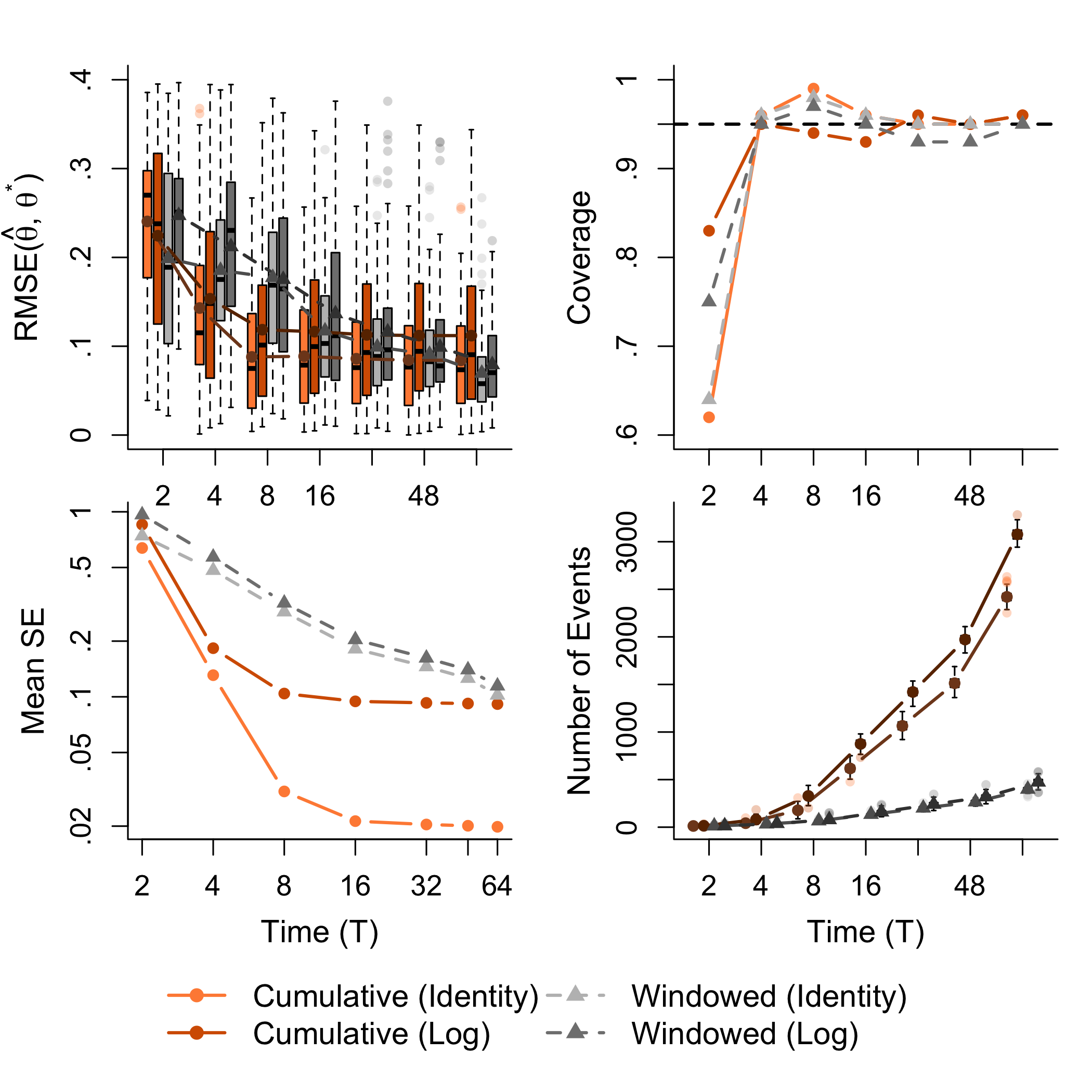}
        \caption{Simulation Study with $T\rightarrow \infty$: Left panels display the estimator RMSE and Mean Standard Error; right panels display empirical coverage probabilities and event counts. }    
                 \label{fig:t}
\end{figure}

\noindent\textbf{Scenario: $T \rightarrow \infty$:}
For the setting with $T\rightarrow \infty$, we utilize the common partner statistic. 
This statistic evaluates the number of common partners for a given dyad $(i,j)\in \dyadset$, being the number of entities $h\in \unitset$ satisfying both $N_{i,h}>0$ and $N_{h,j}>0$. 
Unlike its windowed counterpart, the cumulative specification of this effect lacks temporal ergodicity and ceases to learn anything new at some point in time. 
This phenomenon is illustrated in the lower-left plane of Figure \ref{fig:t} by the flat trajectory of the Mean SE of the cumulative specifications (orange lines), rendering the estimators inconsistent. 
Windowed specifications, on the other hand, learn continuously and stabilize the counting processes, as shown by the steadily decreasing mean standard error and the linearly increasing number of events.   

\section{Discussion}

\hide{
While the asymptotic framework developed in Section~\ref{subsec:general_asymptotics} provides robust inference for a finite-dimensional parameter $\para\in\parspace\subseteq\IR^p$ with a fixed $p$, extending this theory to high-dimensional parametric, semiparametric, and nonparametric settings would be of particular interest.
}

The theory discussed in Section~\ref{subsec:general_asymptotics} covers the case of a finite-dimensional parameter $\para\in\parspace\subseteq\IR^p$ with $p$ fixed. 
Worthwhile generalizations of this framework would be to high-dimensional parametric, semiparametric, and nonparametric settings. 

For high-dimensional models with $p\to\infty$ or even $p>n$, penalized estimation is a classical approach, and inference methods have been discussed in the context of classical counting-process. 
\citet{T09} provides a general discussion of penalized estimation in Cox models, while \citet{YBS21} provide an inferential framework.

If one is primarily interested in a low-dimensional subset of coefficients, another possibility is to consider a semiparametric model incorporating a potentially infinite-dimensional nuisance parameter $\semipara$ 
substituting \eqref{eq:intensity_rem} with:
\beno 
\lambda_{i,j}(t\mid\para,\semipara)&=&g(\para^\top\covs_{i,j}(t)+m(\semipara,\semicovs_{i,j}(t))),
\ee
where $\semicovs_{i,j}$ describe a new set of low-dimensional statistics, and $m$ is a known link function. 
In such a model, the focus remains on estimation of $\para$. In such cases, a partial likelihood approach as described in \cite{MV00} is a promising starting point for inference.

Relaxing the assumption of linear covariate effects introduces nonparametric specifications, such as $\lambda_{i,j}(t\mid\varphi)=\varphi(t,\covs_{i,j}(t))$. 
To bypass the curse of dimensionality plaguing such models, additive or multiplicative intensities could be specified through
\beno 
\lambda_{i,j}(t\mid\varphi)&=&\varphi_0(t)\, \dprod_{q = 1}^p \varphi_q(\covs_{i,j,q}(t)) \text{ and }
\lambda_{i,j}(t\mid\varphi)&=&\varphi_0(t)+ \dsum_{q = 1}^p \varphi_q(\covs_{i,j,1}(t)),
\ee
respectively. 
Robust estimation of the functions $\varphi_0, ..., \varphi_p$ 
could be potentially achieved by adapting smooth backfitting algorithms \citep{HMMMN21, BHMN25} to the REM framework.

Beyond this, the mathematical framework and empirical simulations presented here offer practical guidance for researchers applying REMs to real-world data. 
The lessons learned from this asymptotic theory for the applied model are:
\begin{enumerate}
    \item Relying on raw counts for endogenous statistics, such as the number of prior events of some pair of entities, can lead to explosive counting processes and improper empirical coverage. 
    Applying transformations, such as $\log(\cdot+1)$, is then critical to stabilize the stochastic processes and ensure reliable inference.
    Alternative transformations are bounded functions such as $1/(\cdot+1), \cdot/(\cdot+K),$ or the indicator $\mathbb{I}(\cdot > 0)$ with some $K \in \IN_0$, all available in the $\mathtt{R}$ package $\mathtt{redeem}$ \citep{redeem}. 
    \item In asymptotic regimes where $T\rightarrow \infty$, cumulative statistics lack temporal ergodicity and eventually cease to learn. Employing some form of windowed statistics ensures continuous learning and prevents estimator inconsistency over long time horizons.
    \item When analyzing growing networks where the number of entities approaches infinity, sufficient statistics must be specified to avoid inducing global dependencies. Ensuring that the behavior of disjoint dyads remains asymptotically uncorrelated is necessary for the convergence of the observed information.
\end{enumerate}

\bibliography{references}

\backmatter



\bmhead{Acknowledgements}

We thank the other organizers \textcolor{blue}{and attendees} of the workshop New Trends in Statistical Network Analysis, Carsten Jentsch and Göran Kauermann, where this work originated for fostering a stimulating and productive scientific exchange.


\newpage

\begin{appendices}

\section{Justification for Theorem~\ref{thm:AN}}
\label{app:discussion_HLassumptions}

Theorem~\ref{thm:AN} is a direct consequence of Theorems VI.1.1 and VI.1.2 from \cite{ABGK93} if the following general Assumptions~\ref{assump:ABGK:diff}-\ref{assump:ABGK:consistency} hold true. While these assumptions are general, their exact meaning can be different for different models. We will argue afterwards why our Assumptions~\ref{assump:bounded_stats}-\ref{assump:scaling} imply them. Note that Theorems VI.1.1 and VI.1.2 from \cite{ABGK93} are formulated for general specifications of $\lambda_{i,j}(t\mid\para)$ while we focus here on $\lambda_{i,j}(t\mid\para)=g(\covs_{i,j}(t)^\top\para)$. We therefore reformulate the assumptions in \cite{ABGK93} for this special case when it shortens the formulas.

\begin{assumptionR}[Smoothness of the model]
\label{assump:ABGK:diff}
The function $g$ is continuously differentiable up to third order. Moreover, $\ell_{n,T}(\para)$ may be differentiated with respect to $\para$ up to third order by interchanging differential and integral.
\end{assumptionR}

\begin{assumptionR}[Convergence of observed information]
\label{assump:ABGK:Fisher}
There is an increasing sequence $(a_\sampsize)_{\sampsize\in\IN}\subseteq[0,\infty)$ with $a_\sampsize\to\infty$ such that
$$a_\sampsize^{-2}\dsum_{(i,j) \in \dyadset}\int_\timeset  \left(\partial_{\para_k}\partial_{\para_l}\log\lambda_{i,j}(t;\truepara)\right)^2\lambda_{i,j}(t;\truepara)\dt.$$
converges in probability to a finite constant for all $k,l\in\{1,...,p\}$. There is a positive definite matrix $\Sigma(\truepara)\in\IR^{p\times p}$ such that $a_\sampsize^{-2}F_\sampsize(\truepara)\overset{\IP}{\to}\Sigma(\truepara)$. 
\end{assumptionR}

\begin{assumptionR}[Lindeberg condition]
\label{assump:ABGK:Linderberg}
For all $k\in\{1,...,p\}$, and $\epsilon>0$
\beno 
a_\sampsize^{-2} \dsum_{(i,j) \in \dyadset}\int_\timeset\left(\frac{\partial_{\para_k}\lambda_{i,j}(r\mid \truepara)}{\lambda_{i,j}(r\mid \truepara)}\right)^2\Ind\left(\left|a_\sampsize^{-1}\frac{\partial_{\para_k}\lambda_{i,j}(r\mid \truepara)}{\lambda_{i,j}(r\mid \truepara)}\right|>\epsilon\right)\lambda_{i,j}(r\mid \truepara)\dd r&\overset{\IP}{\to}&0.
\ee
\end{assumptionR}

\begin{assumptionR}[Uniform convergence of observed information]
\label{assump:ABGK:Hesse_uniform}
For any random sequence $\para_\sampsize\overset{\IP}{\to}\truepara$, it holds that $a_\sampsize^{-2}H_\sampsize(\para_\sampsize)\overset{\IP}{\to}-\Sigma(\truepara)$.
\end{assumptionR}

\begin{assumptionR}[Consistency]
\label{assump:ABGK:consistency}
The estimator $\MLE$ is consistent, that is, $\MLE\overset{\IP}{\to}\truepara$.
\end{assumptionR}

Assumptions~\ref{assump:ABGK:Hesse_uniform} and~\ref{assump:ABGK:consistency} are high-level and can be proven under different assumptions for different models. For our case, the following Lemma~\ref{lem:ABGK:bound}, which is also a consequence of Theorems VI.1.1 and VI.1.2 in \cite{ABGK93}, gives more accessible conditions under which Assumptions~\ref{assump:ABGK:Hesse_uniform} and~\ref{assump:ABGK:consistency} hold; but there are other ways of establishing them. They are essentially required to prove that the remainder term in a Taylor expansion converges to zero.

\begin{lemma}
\label{lem:ABGK:bound}
Let Assumptions~\ref{assump:ABGK:diff} and~\ref{assump:ABGK:Fisher} hold. Suppose in addition that there are predictable processes $G_{i,j}:\timeset\to[0,\infty)$ and $H_{i,j}:\timeset\to[0,\infty)$ for all $i,j\in\{1,...,n\}$ and all $n\in\IN$ such that
\begin{align*}
    \max_{k,l,m\in\{1,...,p\}}\sup_{\para\in\paraneighbour}\left|\partial_{\para_k}\partial_{\para_l}\partial_{\para_m}\lambda_{i,j}(t\mid\para)\right|\leq G_{i,j}(t), \\
    \max_{k,l,m\in\{1,...,p\}}\sup_{\para\in\paraneighbour}\left|\partial_{\para_k}\partial_{\para_l}\partial_{\para_m}\log\lambda_{i,j}(t\mid\para)\right|\leq H_{i,j}(t),
\end{align*}
such that
\begin{align}
&a_\sampsize^{-2}\dsum_{(i,j) \in \dyadset} \int_\timeset G_{i,j}(t)\dt,\qquad a_\sampsize^{-2}\dsum_{(i,j) \in \dyadset}\int_\timeset H_{i,j}(t)\lambda_{i,j}(t\mid\truepara)\dt, \nonumber
\end{align}
all converge in probability to finite constants. Finally, suppose that, for any $\epsilon>0$,
$$a_\sampsize^{-2}\dsum_{(i,j) \in \dyadset}\int_\timeset H_{i,j}(t)\Ind\left(a_\sampsize^{-1}\sqrt{H_{i,j}(t)}>\epsilon\right)\lambda_{i,j}(t\mid\truepara)\dt\overset{\IP}{\to}0.$$
Then, Assumptions~\ref{assump:ABGK:Hesse_uniform} and~\ref{assump:ABGK:consistency} hold.
\end{lemma}

We argue next that Assumptions~\ref{assump:ABGK:diff}-\ref{assump:ABGK:consistency} hold under our Assumptions~\ref{assump:bounded_stats}-\ref{assump:scaling}. Note firstly that, by Assumption~\ref{assump:bounded_stats}, the parameter space $\paraspace\subseteq\IR^p$ is bounded, and, hence, it a subset of a compact set $\overline{\paraspace}$.

Assumption~\ref{assump:ABGK:diff} clearly holds as we assume differentiability of $g$ in Assumption~\ref{assump:bounded_stats}. Differentiability of $\ell_{n,T}(\para)$ is then also a direct consequence: Note that the first integral with respect to $N_{i,j}(t)$ can be rewritten as a sum. Therefore, differentiability of it follows by the classical rules of differentiation. The second integral is a parameter integral and can be differentiated under the integral sign if all derivatives of $\lambda_{i,j}(t\mid\para)$ are uniformly bounded on the compact set $\overline{\paraspace}$. This is the case because $\covs_{i,j}(t)$ are uniformly bounded by Assumption~\ref{assump:bounded_stats}, and $g$ is continuously differentiable.

Assumption~\ref{assump:ABGK:Fisher} and Assumption~\ref{assump:Fisher} are identical.

To argue that Assumption~\ref{assump:ABGK:Linderberg} holds, we firstly note that $\lambda_{i,j}(t\mid\para)$ is uniformly bounded away from zero on $\overline{\paraspace}$. More specifically, by Assumption~\ref{assump:bounded_stats}, we have that
$$\inf_{\para\in\overline{\paraspace}}\inf_{n\in\IN}\inf_{t\in\timeset}\min_{i,j=1,...,n}g(\para^\top\covs_{i,j}(t))\geq\epsilon_0>0\text{ almost surely}$$
for a suitable $\epsilon_0>0$ because $\covs_{i,j}(t)$ is uniformly bounded by Assumption~\ref{assump:bounded_stats}. Therefore,
$$\inf_{n\in\IN}\inf_{t\in\timeset}\inf_{\para\in\paraspace}\min_{i,j=1,...,n}\lambda_{i,j}(t\mid\para)\geq\epsilon_0\text{ almost surely.}$$
Furthermore
\begin{align}
&\sup_{n\in\IN}\sup_{t\in\timeset}\max_{i,j=1,...,n}\left|\lambda_{i,j}(t;\truepara)\right|=\left|g(\covs_{i,j}(t)^\top\truepara)\right|\leq M_0\text{ almost surely, and} \nonumber \\
&\sup_{n\in\IN}\sup_{t\in\timeset}\max_{i,j=1,...,n}\left|\partial_{\para_k}\lambda_{i,j}(t;\truepara)\right|=\left|\covs_{i,j,k}(t)g(\covs_{i,j}(t)^\top\truepara)\right|\leq M_1\text{ almost surely} \label{eq:bound_first_derivative}
\end{align}
for suitable constants $M_0,M_1\in(0,\infty)$, by the boundedness assumption in Assumption~\ref{assump:bounded_stats}. We conclude that
\begin{align*}
&a_\sampsize^{-2} \dsum_{(i,j) \in \dyadset} \int_\timeset\left(\frac{\partial_{\para_k}\lambda_{i,j}(r;\truepara)}{\lambda_{i,j}(r;\truepara)}\right)^2\Ind\left(\left|a_\sampsize^{-1}\frac{\partial_{\para_k}\lambda_{i,j}(r;\truepara)}{\lambda_{i,j}(r;\truepara)}\right|>\epsilon\right)\lambda_{i,j}(r;\truepara)\dd r \\
\leq&a_\sampsize^{-2} \dsum_{(i,j) \in \dyadset}\int_\timeset\left(\frac{M_1}{\epsilon_0}\right)^2\Ind\left(a_\sampsize^{-1}\frac{M_1}{\epsilon_0}>\epsilon\right)M_0\dd r \\
=&a_\sampsize^{-2}\frac{n(n-1)}{2}T\frac{M_1^2M_0}{\epsilon_0^2}\Ind\left(\frac{M_1}{\epsilon_0\epsilon}>a_\sampsize\right).
\end{align*}
Since we assume $a_{\sampsize}\to\infty$, the indicator on the right will eventually be equal to zero, and therefore Assumption~\ref{assump:ABGK:Linderberg} holds true.

We argue that Assumptions~\ref{assump:ABGK:Hesse_uniform} and~\ref{assump:ABGK:consistency} hold true by showing the conditions of Lemma~\ref{lem:ABGK:bound}. Note firstly that similarly to \eqref{eq:bound_first_derivative}, we have
\begin{align*}
&\max_{k,l,m\in\{1,...,p\}}\sup_{n\in\IN}\sup_{t\in\timeset}\left|\partial_{\para_k}\partial_{\para_l}\partial_{\para_m}\lambda_{i,j}(t;\truepara)\right|\leq M_3\text{ almost surely, and} \\
&\max_{k,l,m\in\{1,...,p\}}\sup_{n\in\IN}\sup_{t\in\timeset}\left|\partial_{\para_k}\partial_{\para_l}\partial_{\para_m}\log\lambda_{i,j}(t;\truepara)\right|\leq\overline{M}_3\text{ almost surely}
\end{align*}
for suitable constants $M_3,\overline{M}_3\in(0,\infty)$. Then, we may choose $G_{i,j}(t)=M_3$ and $H_{i,j}(t)=\overline{M}_3$ in Lemma~\ref{lem:ABGK:bound}. We conclude that
$$a_{\sampsize}^{-2}\dsum_{(i,j) \in \dyadset}\int_\timeset G_{i,j}(t)\dt=a_{\sampsize}^{-2}\frac{n(n-1)}{2}TM_3=M_3$$
by our choice of $a_\sampsize$ in Assumption~\ref{assump:Fisher}. As a constant, the above expression converges to a finite constant. Moreover, 
$$a_{\sampsize}^{-2}\dsum_{(i,j) \in \dyadset}\int_\timeset H_{i,j}(t)\lambda_{i,j}(t\mid\truepara)\dt=a_{\sampsize}^{-2}\overline{M}_3\dsum_{(i,j) \in \dyadset}\int_\timeset\lambda_{i,j}(t\mid\truepara)\dt,$$
which converges by Assumption~\ref{assump:scaling} to a finite constant. The Lindeberg type condition in Lemma~\ref{lem:ABGK:bound} can be proven by similar arguments as we used for Assumption~\ref{assump:ABGK:Linderberg}.

\subsection{Sketch of Theorem~\ref{thm:AN} proof}
\label{subsec:proof_idea}
We emphasize that Theorem~\ref{thm:AN} is a direct consequence of Theorems~VI.1.1 and~VI.1.2 of \cite{ABGK93}. However, to explain the main ideas, we give here a very short outline of the proof. By differentiability, guaranteed by Assumption~\ref{assump:ABGK:diff}, we have by an application of Taylor's Theorem that
$$0=\partial\ell_\sampsize(\MLE)=U_\sampsize(\truepara)-H_\sampsize(\truepara)(\MLE-\truepara)-(H_\sampsize(\tilde{\para}_\sampsize)-H_\sampsize(\truepara))(\MLE-\truepara),$$
where we abuse notation as follows: By $H_\sampsize(\tilde{\para}_\sampsize)$ we mean a matrix, where the $i$-th row equals the $i$-th row of $H_\sampsize(\para^{(i)})$ with $\para^{(i)}$ lying on the connecting line between $\truepara$ and $\MLE$. In other words, each row of $H_\sampsize(\tilde{\para}_\sampsize)$ has its own intermediate parameter. Rearranging the above equation and multiplication with $a_\sampsize$ yields
\begin{equation}
\label{eq:taylor}
a_\sampsize(\MLE-\truepara)=\left(a_\sampsize^{-2}H_\sampsize(\truepara)+a_\sampsize^{-2}(H_\sampsize(\tilde{\para}_\sampsize)-H_\sampsize(\truepara))\right)^{-1}a_\sampsize^{-1}U_\sampsize(\truepara).
\end{equation}
Assumptions~\ref{assump:ABGK:Hesse_uniform} and~\ref{assump:ABGK:consistency} imply that
$$\left(a_\sampsize^{-2}H_\sampsize(\truepara)+a_\sampsize^{-2}(H_\sampsize(\tilde{\para}_\sampsize)-H_\sampsize(\truepara))\right)^{-1}\overset{\IP}{\to}\Sigma(\truepara)^{-1}.$$
We have argued below Theorem~\ref{thm:AN} that, by Assumptions~\ref{assump:ABGK:Fisher} and~\ref{assump:ABGK:Linderberg}, $a_\sampsize^{-1}U_\sampsize(\truepara)$ converges to a normal distribution $\mathcal{N}(0,\Sigma(\truepara))$. Hence, by \eqref{eq:taylor} we obtain the convergence of $a_\sampsize(\MLE-\truepara)$ to $\mathcal{N}(0,\Sigma(\truepara)^{-1})$.

\subsubsection{Different scaling}
\label{subsubsec:scaling}
Note that $a_\sampsize$ primarily guarantees that $a_\sampsize^{-1}U_\sampsize(\truepara)$ converges to a normal distribution and that $a_\sampsize^{-2}H_\sampsize(\truepara)\to\Sigma(\truepara)$. We would hence typically expect that $a_\sampsize$ equals the square root of $\Var\left(U_{\sampsize,k}(\truepara)\right)$, where $U_{\sampsize,k}$ denotes the $k$-th entry of the vector $U_\sampsize$. A problem arises if different entries of $U_\sampsize(\truepara)$ have different variances. We suppose that $\covs_{i,j}$ are identically distributed but scale differently and that, e.g.,
\begin{align*}
\Var(U_{\sampsize,k}(\truepara))&=\dsum_{(i,j) \in \dyadset}\Var\left(\int_\timeset\partial_{\para_k}\log\lambda_{i,j}(r\mid\truepara)\dM_{i,j}(r)\right) \\
&=\frac{n(n-1)}{2}\int_\timeset\IE\left(\left(\partial_{\para_k}\log\lambda_{i,j}(r\mid\truepara)\right)^2\lambda_{i,j}(r\mid\truepara)\right)\dr \\
&=\frac{Tn(n-1)}{2}v_{\sampsize,k},
\end{align*}
where we have assumed that the variance is the same for all pairs $(i,j)$ and that the integrand in the second line is bounded away from zero. The last equality serves as a definition of $v_{\sampsize,k}$. Let $(v_{\sampsize,k})_{\sampsize}$ denote sequences that potentially have different growth rates for different $k$. 
Let $\mathbf{D}_\sampsize\in\IR^{p\times p}$ be a sequence of diagonal matrices with diagonal entries $(\sqrt{v_{\sampsize,k}})_{k=1}^p$. 
It is then plausible that for $a_\sampsize=\sqrt{Tn(n-1)/2}$, we have that $a_\sampsize^{-1}\mathbf{D}_\sampsize^{-1} U_\sampsize(\truepara)$ converges to a normal distribution. It follows, in turn, that $a_\sampsize^{-2}\mathbf{D}_\sampsize^{-1}F_\sampsize(\truepara)\mathbf{D}_\sampsize^{-1}$ plausibly converges to some constant matrix $\Sigma(\truepara)$. In a Cox model, one can directly conclude that $a_\sampsize^{-2}\mathbf{D}_\sampsize^{-1}H_\sampsize(\truepara)\mathbf{D}_\sampsize^{-1}$ converges to $\Sigma(\truepara)$. In other models, an equivalent to the first part of Assumption~\ref{assump:ABGK:Fisher} is required. Moreover, we have to prove that this convergence holds uniformly as in Assumption~\ref{assump:ABGK:Hesse_uniform}. If we can show this, we can multiply \eqref{eq:taylor} from the left with $\mathbf{D}_\sampsize$ and obtain
\begin{align}
a_\sampsize\mathbf{D}_\sampsize(\MLE-\truepara)&=\left(a_\sampsize^{-2}\mathbf{D}_\sampsize^{-1}H_\sampsize(\truepara)\mathbf{D}_\sampsize^{-1}+a_\sampsize^{-2}\mathbf{D}_\sampsize^{-1}(H_\sampsize(\tilde{\para}_\sampsize)-H_\sampsize(\truepara))\mathbf{D}_\sampsize^{-1}\right)^{-1} \nonumber \\
&\qquad\times a_\sampsize^{-1}\mathbf{D}_\sampsize^{-1}U_\sampsize(\truepara). \nonumber
\end{align}
If we believe the assumptions that we have just discussed, we can now conclude as in Section~\ref{subsec:proof_idea} that
$$a_\sampsize\mathbf{D}_\sampsize\left(\MLE-\truepara\right)\overset{d}{\to}\mathcal{N}\left(0,\Sigma(\truepara)^{-1}\right).$$

\section{More details to Section~\ref{subsec:Cox}}
\label{app:Cox}

To give more details on the heuristic discussion in Section~\ref{subsec:Cox}, note that
\begin{align*}
&\textrm{Cov}\left(\int_\timeset F_{i,j}(t)\dt,\int_\timeset F_{k,l}(t)\dt\right) \\
&\qquad=\int_\timeset \int_\timeset\underbrace{\IE\left(F_{i,j}(t)F_{k,l}(t)\right)}_{=:\phi_{ij,kl}(t,s)}\dt\ds-\left(\int_\timeset\underbrace{\IE\left(F_{i,j}(t)\right)}_{=:\psi_{ik}(t)}\dt\right)^2.
\end{align*}
Applying the above in \eqref{eq:two_overlap} yields
\begin{align*}
\eqref[s]{eq:two_overlap} 
=&\frac{n(n-1)}{2\epsilon^2a_\sampsize^4}\textrm{Var}\left(\int_\timeset F_{1,2}(t)\dt\right) \\
=&\frac{2}{\epsilon^2\cdot n(n-1)}\left(\frac{1}{T^2}\int_\timeset \int_\timeset \phi_{12,12}(t,s)\dt\ds-\left(\frac{1}{T}\int_\timeset \Psi_{12}(t)\dt\right)^2\right),
\end{align*}
which converges to zero, when assuming that $n\to\infty$ and
\begin{equation}
\label{eq:bounded_var}
\frac{1}{T^2}\int_\timeset\int_\timeset\IE\left(F_{1,2}(t)F_{1,2}(s)\right)\dt\ds=O(1)\text{ and }\frac{1}{T}\int_\timeset\IE\left(F_{1,2}(t)\right)\dt=O(1).
\end{equation}
These assumptions seem plausible for $T\to\infty$ or $T$ fixed, and many choices of sufficient statistics.
For \eqref{eq:one_overlap}, we obtain in a similar fashion that
\begin{align*}
\eqref[s]{eq:one_overlap}=&\frac{2(n-2)}{\epsilon^2n(n-1)}\left(\frac{1}{T^2}\int_\timeset\int_\timeset\phi_{12,13}(t,s)\dt\ds-\left(\frac{1}{T}\int_\timeset\psi_{12}(t)\dt\right)^2\right).
\end{align*}
The above expression converges to zero when $n\to\infty$ and under a boundedness assumption on the integrals similar to \eqref{eq:bounded_var}.

Finally, \eqref{eq:no_overlap} can be rewritten by the same logic to
\begin{align*}
\eqref[s]{eq:no_overlap} =&\frac{(n-2)(n-3)}{\epsilon^2n(n-1)}\left(\frac{1}{T^2}\int_\timeset\int_\timeset\phi_{12,34}(t,s)\dt\ds-\left(\frac{1}{T}\int_\timeset\psi_{12}(t)\dt\right)^2\right)
\end{align*}
The above converges to zero for $n\to\infty$ when one assumes that the behavior on disjoint edges becomes asymptotically uncorrelated.
Specifically, we require that (recall that $\psi_{12}(s)=\psi_{34}(t)$)
\begin{align*}
\phi_{12,13}(t,s)=&\IE(F_{1,2}(t)F_{3,4}(s))\approx\IE(F_{1,2}(t))\IE(F_{3,4}(s))=\psi_{12}(t)\psi_{12}(s)
\end{align*}
by concretely assuming that
$$\frac{1}{T^2}\int_\timeset\int_\timeset\phi_{12,34}(t,s)\dt\ds-\left(\frac{1}{T}\int_\timeset\psi_{12}(t)\dt\right)^2\to0.$$
To give a heuristic motivation for the last assumption, consider the following set-up. Suppose that there is an underlying, potentially unknown, sparse network $G$ that contains all dyads $(i,j)\in\dyadset$ as vertices.
Let $d((i,j),(k,l))$ denote the length of the shortest path between dyads $(i,j)$ and $(k,l)$ within this network.
We assume that the amount of dependence between $\covs_{i,j}$ and $\covs_{k,l}$ decreases when $d((i,j),(k,l))$ increases.
Then, the amount of dependence of $F_{i,j}$ and $F_{k,l}$ is also decaying when distance between $(i,j)$ and $(k,l)$ in the network $G$ increases.
Assume finally that $d((i,j),(k,l))\geq 1$ when $\{i,j\}\cap\{k,l\}=\emptyset$.
In such scenarios it is plausible that
$$\IE(F_{1,2}(t)F_{3,4}(s)\mid d((1,2),(3,4))=k)\approx\IE(F_{1,2}(t))\IE(F_{3,4}(s)),$$
when $k$ is large.
Moreover, in sparse networks it is much more likely to observe a large $d((1,2),(3,4))$.
Hence, the large distance expectations play an increasingly important role in
\begin{align*}
&\IE(F_{1,2}(t)F_{3,4}(s)) \\
=&\sum_{d=1}^{\infty}\IE\left(F_{1,2}(t)F_{3,4}(s)\mid d((1,2),(3,4))=d\right)\IP\left(d((1,2),(3,4))=k\right),
\end{align*}
and, hence, we may expect asymptotically that disjoint edges are uncorrelated.

\section{Simulation Study Setup}
\label{sec:sim_setup}
Table \ref{tab:sim} summarizes the configuration of the simulation study designed to evaluate the finite-sample behavior of the estimator under both spatial ($n \to \infty$) and temporal ($T \to \infty$) asymptotic regimes. 
The table details the chosen sufficient statistics, observation grids, and true parameter vectors across different functional transformations.

\begin{table}[t!]
\centering
\caption{Configuration of simulation studies. 
The number of independent replications is set to $100$ for all scenarios. 
For the windowed statistics, the window is fixed at 0.5 for Simulation Study 1 and at 1 for Simulation Study 2.}
\label{tab:sim}
\begin{tabular*}{\textwidth}{@{\extracolsep{\fill}} l c c }
\toprule
& \textbf{Simulation 1} & \textbf{Simulation 2} \\
\midrule
\textbf{Regime}      & $n \to \infty$ & $T \to \infty$ \\
\textbf{Statistics}   & (Intercept, Inertia) & (Intercept, Common Partner) \\
\textbf{Fixed} & $T=1$ & $n=10$ \\
\textbf{Grid} & $n \in \{35, 40, \dots, 70\}$ & $T \in \{2, 5, \dots, 60\}$ \\
\midrule
\multicolumn{3}{c}{\textbf{True Parameters}} \\
\midrule
Raw           & $(-2.95, 1.2)$ & $(-2.0, 0.25)$ \\
Log           & $(-1.5, 1.2)$  & $(-2.0, 1.0)$ \\
Windowed Raw  & $(-2.5, 1.2)$ & $(-2.0, 0.45)$ \\
Windowed Log  & $(-1.5, 1.2)$ & $(-2.0, 1.2)$ 
\end{tabular*}
\end{table}

\end{appendices}
%


\end{document}